\documentclass[journal]{IEEEtran}
\pdfoutput=1
\usepackage{amsmath}
\usepackage{amssymb}
\usepackage{cite}
\usepackage{algorithm}
\usepackage{algpseudocode}
\usepackage{subfig}
\usepackage{graphicx}
\usepackage{epstopdf}
\usepackage{color}
\usepackage{enumitem}
\usepackage{microtype}
\usepackage{setspace}
\usepackage{makecell}
\usepackage{cases}
\usepackage{ntheorem}
\usepackage{multirow}
\usepackage{pifont}
\usepackage{diagbox}
\usepackage{newunicodechar}

\newunicodechar{，}{,}

\begin{document}
	
	\title{Blockchain-Based Identity Authentication\\Oriented to Multi-Cluster UAV Networking}
	\author{Zesong Dong, Wei Tong*, Zhiwei Zhang, Jian Li, Weidong Yang, and Yulong Shen
		\thanks{Wei Tong, Zesong Dong, Jian Li, and Weidong Yang are with the Hangzhou Institute of Technology, Xidian University, Hangzhou, China. Zhiwei Zhang and Yulong Shen are with the School of Computer Science and Technology, Xidian University, Xi'an, China. Wei Tong is also with the Engineering Research Center of Blockchain Technology Application and Evaluation, Ministry of Education, Xi'an, China. All authors are also with Shaanxi Key Laboratory of Network and System Security, Xi'an, China.}
		\thanks{This work is supported in part by National Key Research and Development Program	of China (Program No. 2020YFB1005500), in part by National Natural Science Foundation of China under Grant No. 62220106004, 62172141, and 61972310, in part by Major Research Plan of the National Natural Science Foundation of China under Grant No. 92267204, in part by ''Pioneer'' and ''Leading Goose'' R\&D Program of Zhejiang (Program No. 2023C04038), in part by Natural Science Foundation of Henan under Grant No. 222300420004, in part by Key Research and Development Program of Shaanxi under Grant No. 2022KXJ-093 and Grant No. 2021ZDLGY07-05, in part by Innovation Capability Support Program of Shaanxi (Program No. 2023-CX-TD-02), in part by Key R\&D Program of Shandong Province, China Grant No.2023CXPT056, in part by Jinan ''20 New Colleges and Universities'' Introduction and Innovation Team (No. 2021GXRC064), and in part by Proof of Concept Foundation of Xidian University Hangzhou Institute of Technology under Grant No. 20107230024.}
		\thanks{*Corresponding author (email: tongwei@xidian.edu.cn).}
	}
	
	\maketitle
	
	\begin{abstract}
		\textls[-25]{Unmanned Aerial Vehicle (UAV) networking is increasingly used in field environments such as power inspection, agricultural plant protection, and emergency rescue. To guarantee UAV networking security, UAV identity authentication attracts wide attention, especially in the field environment without perfect infrastructure. Some blockchain-based UAV identity authentication solutions are proposed to establish decentralized and trusted authentication systems without relying on infrastructure. However, these solutions do not support disconnected UAV reconnection or even disband a cluster directly after its head UAV disconnection, which compromises cluster robustness and task result integrity. In this paper, we propose a blockchain-based identity authentication solution oriented to multi-cluster UAV networking with a UAV disconnection mechanism and a task result backup mechanism. Specifically, we build a blockchain maintained by head UAVs of all clusters, managing identity information to guarantee the security of decentralized identity management. The UAV disconnection mechanism permits a verified distributed UAV reconnection to ensure the robustness of the UAV cluster, and on this basis, the task result backup mechanism ensures the integrity of the task results stored in a cluster even any UAV disconnection. Finally, extensive experimental results prove the superiority of our solutions in terms of robustness, integrity, delay, and energy consumption.}
	\end{abstract}
	\IEEEpeerreviewmaketitle
\vspace{-0.8em}
	\section{Introduction}
	\label{sec_intro}
	Nowadays, outdoor work is often done using Unmanned Aerial Vehicle (UAV) networking to reduce costs and personal dangers\cite{chen2022dronetalk}. For example, UAVs can complete precise rescue positioning, rapid disaster assessment, and efficient material delivery in disaster areas that are difficult for humans to enter \cite{savkin2020navigation}. However, in such disaster areas, severe infrastructure damage leads to security risks (e.g., Sybil attacks and key leakage) for traditional UAV identity authentication relying on infrastructure, easily causing safety issues, like UAV disconnection and crashes. The robustness of UAV clusters through identity authentication is currently a research hotspot\cite{tanveer2021ramp}. Decentralized blockchain technology is a promising selection, and it can establish trustworthy identity management and authentication without a centralized authentication authority\cite{rahmani2022olsr+}.
	
	\textls[-5]{\textbf{Motivation.} Some blockchain-based UAV identity authentication works have been proposed \cite{tan2020blockchain,wang2021blockchain,zhang2022hybrid,xue2022distributed,tong2022chchain,tong2022blockchain,10198563}, but they cannot address the UAV reconnection issue. Indeed, they directly discard disconnected UAVs and even disband clusters with a disconnected head UAV for security. On one hand, as UAVs continuously disconnect, the number of UAVs that can normally execute tasks decreases, causing an increasing workload for the remaining UAVs. The cluster even disbands directly when the head UAV disconnects, leading to task execution to be interrupted. On the other hand, due to no infrastructure relied on, task results are stored within UAVs. The UAV disconnection also results in the loss of task results from the cluster. The occurrence of these situations compromises the robustness of the UAV cluster and the integrity of the task results stored in the cluster\cite{wang2021blockchain}.}
	
	\textbf{Contribution.} To address these issues, this paper proposes a blockchain-based identity authentication solution oriented to multi-cluster UAV networking. Our solution includes a UAV reconnection mechanism to ensure cluster robustness and a task result backup mechanism to ensure task result integrity. The contributions of this paper are as follows.
	\begin{itemize}
		\item \textbf{A blockchain-based UAV identity authentication model} is constructed, and \textbf{the identity management blockchain} is maintained by the head UAVs of all clusters. All authentication information is stored on-chain to support UAV reconnection and task result backup.
		\item On this basis, we propose \textbf{a UAV reconnection mechanism} containing the UAV disconnection, and UAV reconnection. It can guarantee cluster security during UAV disconnection and reconnection processes via event-driven key updates, ultimately ensuring cluster robustness. We further design\textbf{ a redundant task result backup mechanism} with the double backup of task results by the head UAV and member UAVs. Regardless of the head UAV or member UAV disconnection, the task results will not be completely lost, thus ensuring task result integrity.
		\textls[-25]{\item Extensive experiments are conducted to demonstrate the excellent robustness of the UAV cluster and the integrity of task results. In addition, we analyze delay and energy consumption performance to prove the feasibility of our solution. The delay of UAV reconnection identity authentication of our solution is less than 50\% of that of the existing solution.}
	\end{itemize}
	The remainder of this paper is organized as follows. The system model and scheme of the blockchain-based identity authentication are first presented in Section \ref{sec_so} and Section \ref{sec_ims} respectively. Section \ref{sec_sa&ee} then shows experimental results, and finally, Section \ref{sec_con} concludes this paper.

	\vspace{-0.8em}
	\section{System Overview}
	\label{sec_so}
	
	\subsection{System Model}
	\label{ssec_sm}
\begin{figure}[t]
		\centering
		\includegraphics[width= 0.9\linewidth]{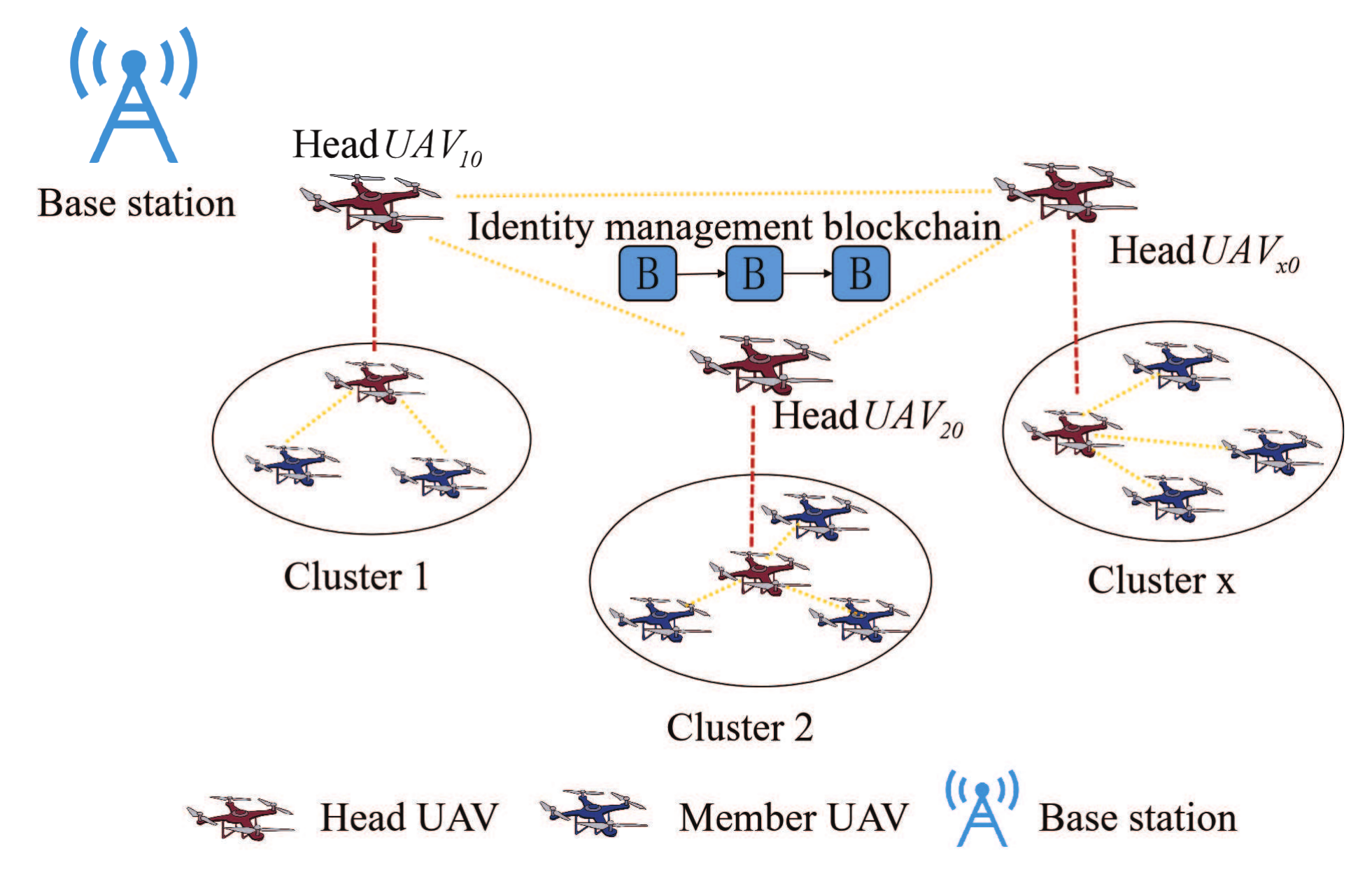}
		\caption{System model.}
		\label{fig:enter-label}
		\vspace{-1em}
	\end{figure}

	As shown in Fig. \ref{fig:enter-label}, the system model consists of four entities, including base station (BS), head UAVs, member UAVs, and identity management blockchain (IDBC). The detailed descriptions of each entity are as follows.
	
	\begin{itemize}
		\item BS manages all UAV clusters and performs initialization operations before UAVs take off, including registration of UAVs, generation of the genesis block, and initial task deployment of all UAV clusters. BS stores assigned key pairs and trained physical layer feature fingerprints on IDBC. It is worth noting that BS participates in initialization operations on the ground rather than identity management of UAVs in the air.	
		\textls[-15]{\item Head UAVs act as hubs within clusters and only manage identity and task information without task execution. Specifically, they maintain IDBC recording key and head UAV candidate update and verify reconnected UAVs. They also distribute tasks in plaintext and collect task results for ciphertext.}		
		\item Member UAVs receive identity update information passively and actively execute tasks, return task results to the head UAVs, and store task results for ciphertext locally. In addition, an original member UAV can be elected as a new head UAV after the preview head UAV disconnection.		
		\textls[-15]{\item There is only one blockchain in the multi-cluster UAV networking \cite{8761147} used to record global information, named IDBC. IDBC is initialized by BS and maintained jointly by all head UAVs. It stores all public information of UAVs, including initialization information, physical layer information, head UAV candidate update records, UAV state records, and a complete task list for all UAV clusters. Besides, IDBC realizes legality detection of head UAV candidate and UAV state change via on-chain smart contracts.} 		
	\end{itemize}
	
	\subsection{Threat model}
	\label{ssec_tm}
	BS is completely honest during the initialization phase, and its operations are completed in a secure environment. Head UAVs can allocate tasks and maintain IDBC honestly, and once each of them disconnects and then reconnects, it can only act as a member UAV. Member UAVs are semi-trusted and obey rules as long as they are not hijacked. Besides, a member UAV and a head UAV in the same cluster cannot disconnect simultaneously, and the number of disconnected head UAVs cannot exceed half of the original total number of head UAVs.
	
\begin{figure}[t]
		\centering
		\subfloat[Attacks through Adversaries' UAVs]{\includegraphics[height=0.9in]{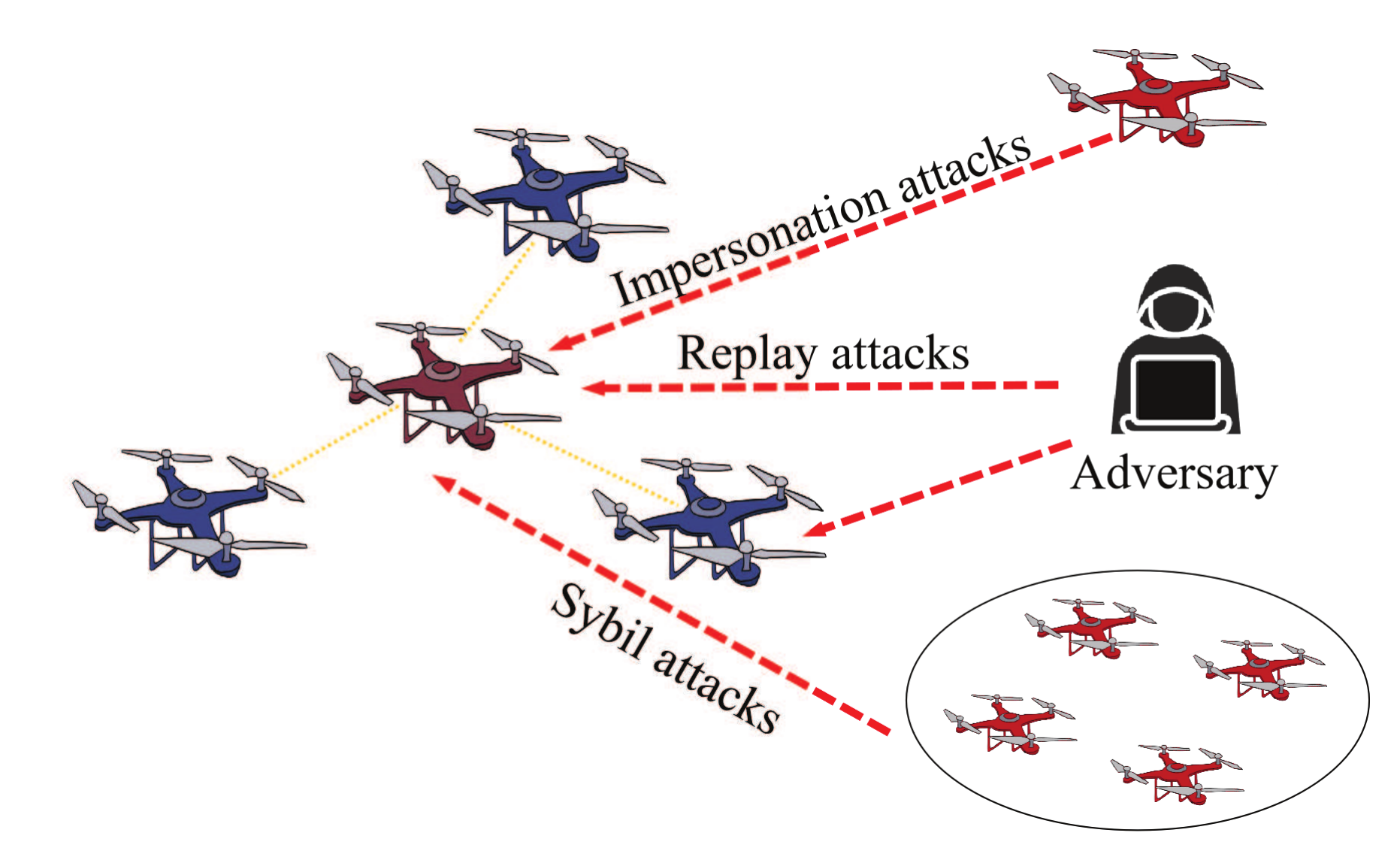}
			\label{Attacks through Adversaries' UAVs.}}
		\subfloat[Attacks through hijacked UAVs of the cluster]{\includegraphics[height=0.9in]{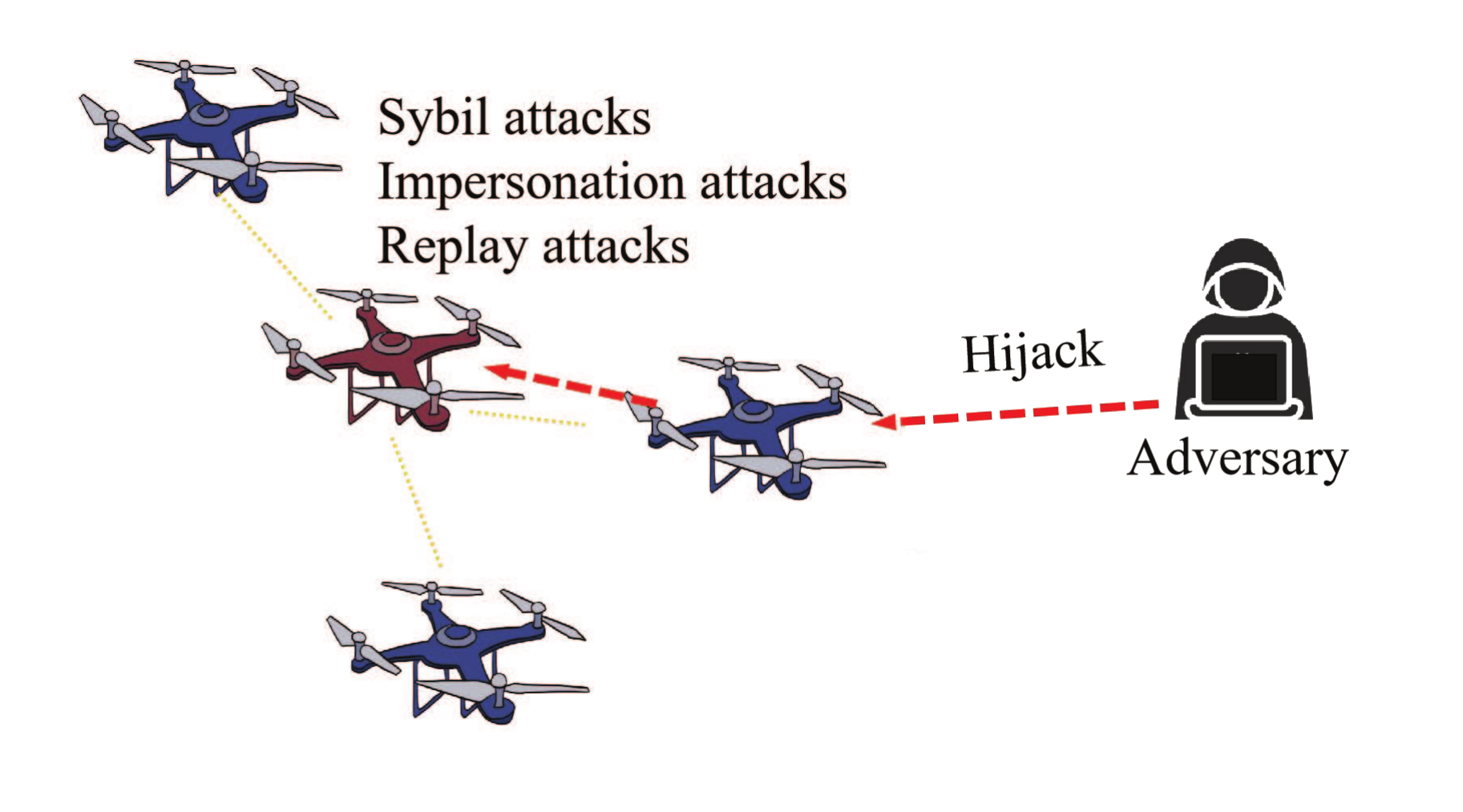}
			\label{Attacks through hijacked UAVs of the cluster.}}
		\caption{Threat Model.}
		\label{fig_dtme}
	\end{figure}

	\textls[-15]{Adversaries can carry out two types of malicious attacks. On one hand, adversaries directly attack clusters through their own UAVs, as shown in Fig. \ref{Attacks through Adversaries' UAVs.}. On the other hand, adversaries attack through hijacked UAVs within clusters, as shown in Fig. \ref{Attacks through hijacked UAVs of the cluster.}.}
	
	In addition, when head UAVs or member UAVs are disconnected due to harsh environments, some task data is lost resulting in incomplete task results.
	\vspace{-0.8em}
	\subsection{System Goals}
	\label{ssec_sg}
	\begin{itemize}
		\item {\textbf{Cluster robustness}. The UAV cluster can continue to perform tasks without disbanding, even if the head UAV or member UAV is disconnected. Besides, after security authentication, the disconnected UAV can reconnect to the previous cluster for work.}
		
		\item \textbf{Task result integrity}. The task results stored by the UAV cluster will not be missing due to the disconnection or being attacked of the head UAV or member UAV.
	\end{itemize}
	\vspace{-0.5em}
	\section{Blockchain-Based Identity Management Scheme}
	\label{sec_ims}
	\textls[-25]{This section proposes a blockchain-based identity management scheme for UAV networking, including initialization, UAV disconnection, UAV reconnection, and task assignment \& delivery of task results. The whole symbols used in this paper are shown in Table \ref{tab_comp}.}
	\begin{table}[t]
		\centering
		\caption{Symbols and explanation.}
		\renewcommand{\arraystretch}{1.1}
		\small
		\label{tab_comp}
		\begin{tabular}{c|l}
			\hline
			{\textbf{Symbol}}&{\textbf{Explanation}}	\\
			\hline
			$E(\mu),G,P,q$ & {Parameters of non singular elliptic curves} \\
			\hline
			$H_1$ & {Hash function for head UAV selections} \\
			\hline
			$H_2$ & {Hash function for packet transmission} \\
			\hline
			$m$ & {Number of UAV clusters} \\
			\hline
			$n$ & {Number of UAVs in a cluster} \\
			\hline
			$C_x$ & {The $x$-th cluster} \\
			\hline
			$UAV_{xy}$ & {The $y$-th UAV in the $x$-th cluster} \\
			\hline
			${STATUS}$ & {UAV's status in LML} \\
			\hline
			${\mathbb{Z}_q^* }$ & {Multiplicative group of integers modulo $q$} \\
			\hline
		\end{tabular}
		\vspace{-0.5em}
	\end{table}
	\vspace{-0.8em}
	\subsection{Initialization}
	\label{ssec_init}
	Before flying, BS completes global initialization, UAV registration, and blockchain initialization. The whole processes are executed in a secure environment.
	
	\subsubsection{Global Initialization}
	\label{sssec_ginit}
	BS selects a non-singular elliptic curve $E(\mu)$, where $\mu$ is a large prime number, and the points on it form a cyclic additive group $G$. $P$ is a generator of $G$ with order $q$. Then, BS randomly selects $SK_B \in \mathbb{Z}_q^* $ as its private key, and the corresponding public key is computed by $PK_B = SK_B \cdot P$. Finally, two hash functions, $H_1$ and $H_2$, are selected. In the end, BS keeps $SK_B$ confidential and publishes the system parameters $(\mu,q,P,PK_B,H_1,H_2)$.
	
	\subsubsection{UAV Registration}
	\label{sssec_uavr}
	\textls[-15]{BS authenticates the identities of UAVs and assigns the authenticated UAVs into corresponding clusters. After assignment, a UAV is selected as the head UAVs and others are member UAVs. The initial head UAV is randomly selected from UAVs with the same performance by human.}
	
	\textls[-35]{We assume that the total number of clusters is $m$. $C_x$ represents the $x$-th cluster, $\text{UAV}_{xy}$ represents the $y$-th UAV in the cluster $C_x$, and $\text{UAV}_{x0}$ represents the head UAV of the cluster $C_x$. For each cluster, taking  $C_x$ as an example, BS distributes the public/private key pairs to all UAVs and generates the initial latest member list  $LML_x = \{ \text{PK}_{x0}:\text{$STATUS$}, \text{PK}_{x1}:\text{$STATUS$}, \ldots,  \text{PK}_{xn}:\text{$STATUS$} \}$, where $n$ denotes the number of UAVs in cluster $C_x$ and $STATUS$ denotes the UAV's status which containing $Native$, $Disconnected$, $Marked$, $Reconnected$. Each UAV stores its public/private key pair and $LML_x$, and secretly keeps its private key. At initialization, all UAVs' status is $Native$ (original member without reconnection).}
	
	\textls[-50]{BS also sets up a white-gray list based physical layer features for each cluster according to LML and each UAV's feature fingerprint trained by physical layer authentication models \cite{10012578,liu2023exploiting}. UAVs in different lists have different capabilities. Specifically, the task results submitted by UAVs in the white list are usually credible without additional verification, while UAVs in the gray list cannot. At initialization, all UAVs store the white-gray list of its cluster and are also on the white list.}
	
	After all clusters have completed initialization, BS generates the initial cluster head list$ = \{ C_1: \text{PK}_{10},  C_2: \text{PK}_{20},  \ldots  , C_n: \text{PK}_{n0} \}$, containing the head UAVs' public keys. The initial cluster header list (CHL) is stored in head UAVs for blockchain leader election. BS sets up two hash functions that $H_1$ is used for member UAVs to participate in the head UAV selection, and $H_2$ is used for data packet transmission between UAVs. In addition, BS initializes the task lists for each cluster, allocating the complete task list that each cluster needs to accomplish.
	
	\subsubsection{Blockchain Initialization}
	\label{sssec_bcinit}
	\textls[-15]{All head UAVs also store a copy of the blockchain which records the candidate UAV sequences for all clusters, physical layer identity authentication models, physical layer features  fingerprint for all UAVs, and completed task lists for all clusters. Also, on-chain smart contracts are used to respond to the head UAV's operations (e.g., UAV disconnection, UAV reconnection, and task execution). BS initializes blockchain with initial transactions, whose specific format is as follows.}
	\begin{flalign}
		\label{equ_initsf1}
		\scriptsize	
		header = \left\{\begin{array}{lll}
			Input address = NULL, \\
			Cluster index = C_x, \\
			Generator = BS.
		\end{array}\right.&&
	\end{flalign}
	\begin{flalign}
		\label{equ_initsf2}
		\scriptsize
		body = \left\{\begin{array}{lll}
			Type = initialization, \\
			Payload = PK_{xy},T,Sign_{SK_B}\{H ( PK_{xy}, T) \},
		\end{array}\right.&&
	\end{flalign}
	
	\textls[-15]{where $Sign_{SK_B}$ refers to the signature generated with BS's private key, $Generator$ points to the role in which the transaction occurs, not the role that submits the transaction. $Input$ $address$ points to the hash value of the previous transaction on the $Generator$. Since this is initialization, this value is $NULL$.}
	\vspace{-1em}
	\subsection{UAV Disconnection}
	\label{ssec_uavdis}
	\subsubsection{Member UAV Disconnection}
	\label{sssec_muavdis}
\begin{figure}[t]
		\centering
		\includegraphics[width=1\linewidth]{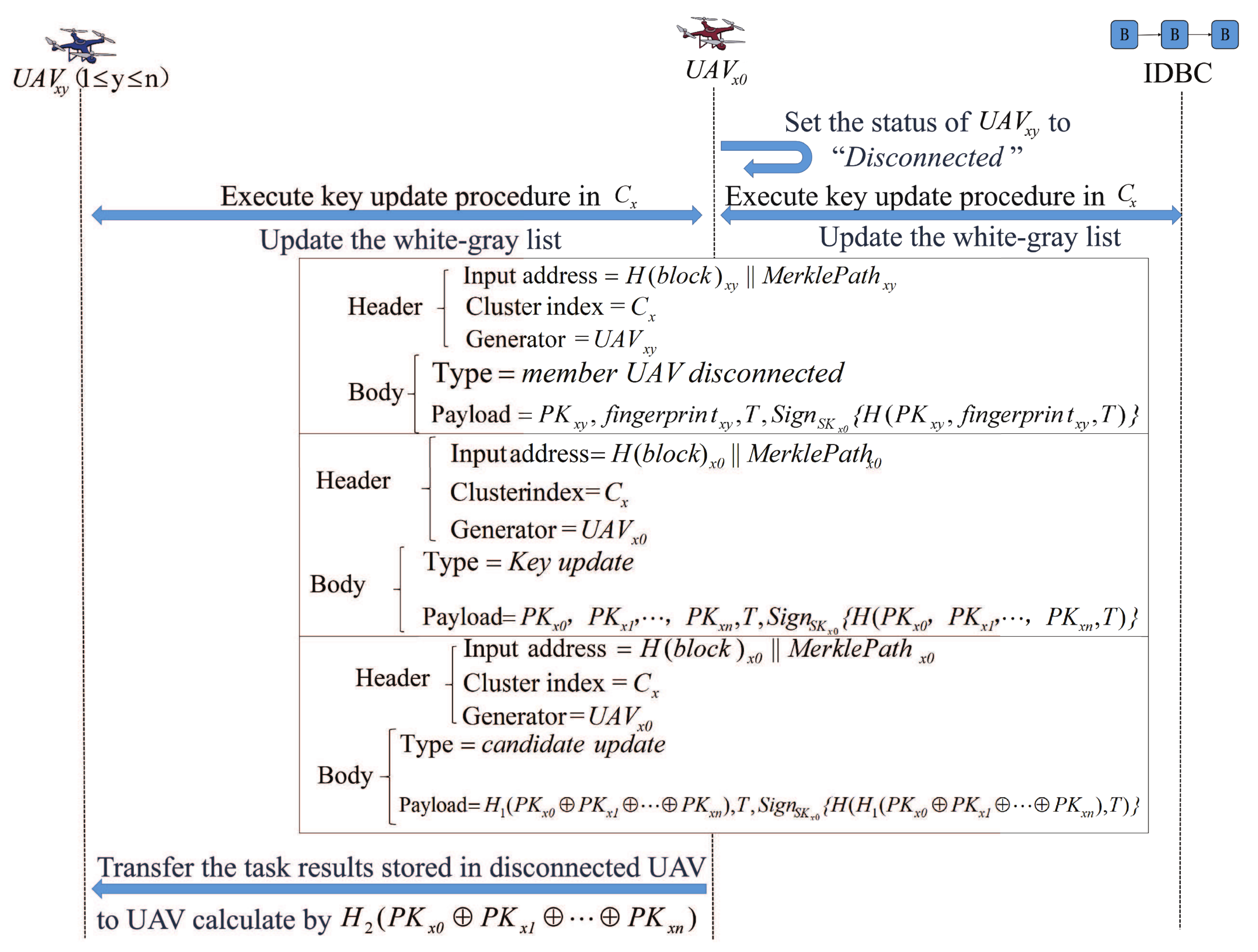}
		\caption{Member UAV disconnection.}
		\label{fig:Member UAV disconnection.}
		\vspace{-1em}
	\end{figure}

	\textls[-5]{Even if there is no need for communication related to tasks, member UAVs must regularly send "Hello" messages to the head UAV to indicate their presence. Suppose the head UAV $UAV_{x0}$ has not detected a signal from the member UAV $UAV_{xy}$ for a prolonged period. In this case, the head UAV assumes that the member UAV $UAV_{xy}$ has been disconnected and thus initiates the disconnection procedure for the disconnected member UAV as shown in Fig. \ref{fig:Member UAV disconnection.}}.
	
	\begin{enumerate}
		\item \textbf{Find a disconnected member UAV.} If the head UAV has not received a ``Hello" signal from a member UAV and also no response to the head UAV's active request, the head UAV determines that the member UAV is disconnected and sets its status to ``$Disconnected$" in LML.
		
		\item \textbf{Execute the key update procedure and update the white-gray list.} The head UAV and IDBC execute the key update procedure to ensure cluster security. In this procedure, the head UAV initiates by creating and sharing a new key with member UAVs. These UAVs validate and share their keys. The head UAV consolidates these into a transaction on the IDBC, which, after approval, is sent back to the head UAV. The process concludes with the head UAV broadcasting updates to the cluster and generating both "key update" and "candidate update" transactions. In addition, the head UAV generates and issues an additional UAV disconnection transaction and updates the white-gray list to add the disconnected UAV. The member UAV disconnection transaction format is as follows.
		\begin{flalign}
			\label{equ_memdc1}
			\scriptsize
			header = \left\{\begin{array}{lll}
				Input address = H(block)_{xy} || MerklePath_{xy}, \\ 
				Cluster index = C_x, \\ 
				Generator = UAV_{xy}.
			\end{array}\right.&&
		\end{flalign}		
		\begin{flalign}
			\label{equ_memdc2}
			\scriptsize
			body = \left\{\begin{array}{ll}
				Type = member\ UAV\ disconnected,& \\ 
				Payload = PK_{xy},fingerprint_{xy},T,\\
				\ \ \ \ \ \ \ \ \ \ \ \ \ Sign_{SK_{x0}}\{H (PK_{xy},fingerprint_{xy}, T) \},
			\end{array}\right.&&
		\end{flalign}			
		
		\textls[-25]{Inspired by Ref. \cite{zamani2018rapidchain}, member UAVs within cluster $C_x$ calculate the candidate head UAV sequence number using \( H_1 (PK_{x0} \oplus PK_{x1}\oplus \ldots \oplus PK{xn}) \). Only UAVs with ``$Native$" status can become candidate UAVs. Since the same LML is used as input, the computation results should theoretically be consistent across all UAVs. We assume the computation identifies member UAV $UAV_{xk}$ as the candidate.}
		
		\item \textls[-15]{\textbf{Copy task results.} After updating and broadcasting LML, the head UAV calculates the formula \( H_2 (PK_{x0} \oplus PK_{x1}\oplus \ldots \oplus PK{xn}) \) to obtain the serial number of a member UAV, and also sends the task results originally stored in the disconnected UAV to this member UAV for copying.}
		
		
		
	\end{enumerate}
	
	\subsubsection{Head UAV Disconnection}
	\label{key}
\begin{figure}[t]
		\centering
		\includegraphics[width=1\linewidth]{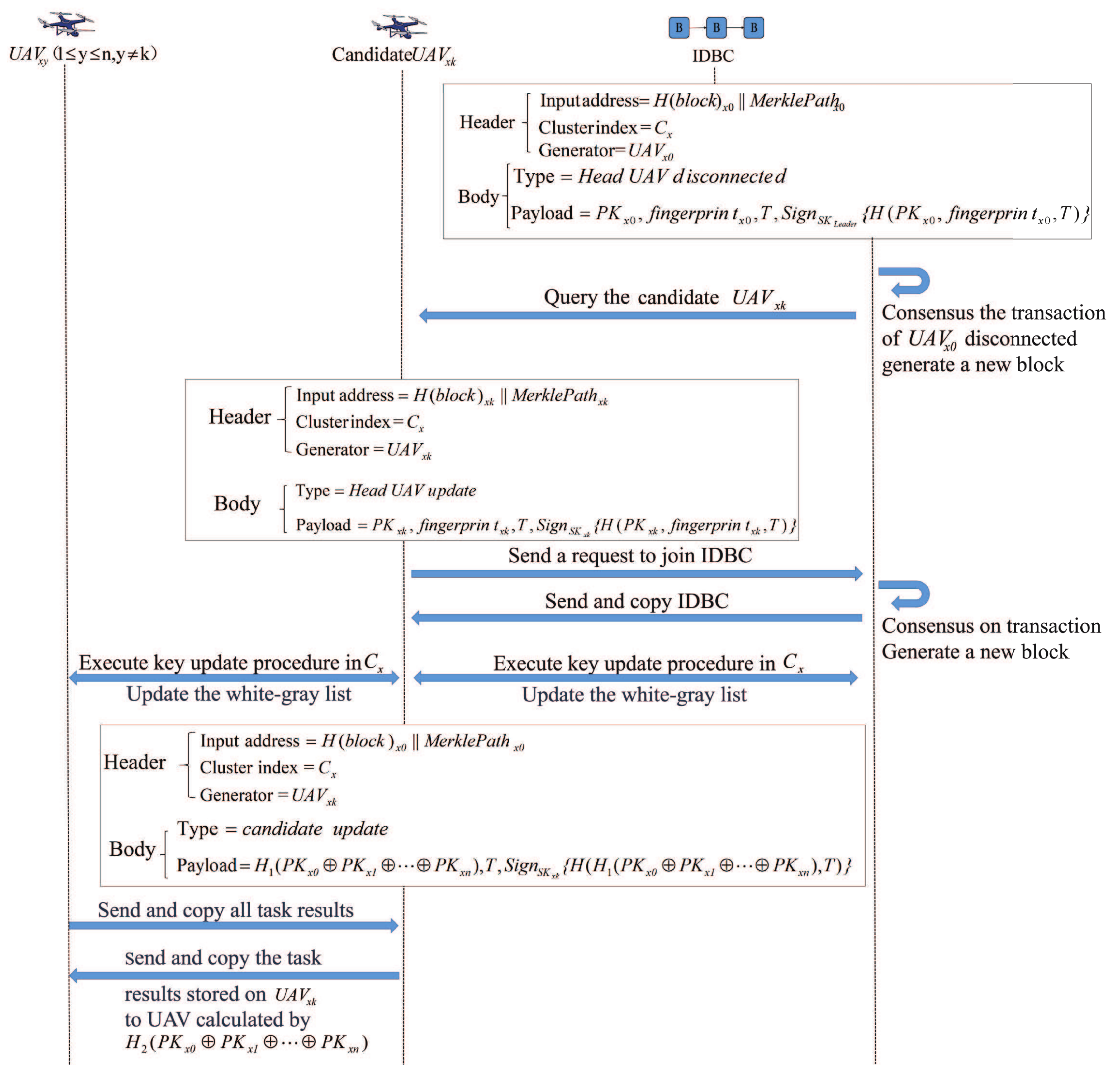}
		\caption{Head UAV disconnection.}
		\label{fig:Head UAV disconnection}
		\vspace{-0.7em}
	\end{figure}

	In addition to member UAV disconnection, the head UAV can also disconnects. Fig. \ref{fig:Head UAV disconnection} shows the head UAV disconnection workflow with seven steps. Suppose the head UAV $UAV_{x0}$ disconnects and the member UAV $UAV_{xk}$ is a candidate head UAV.
	\begin{enumerate}
		\item \textbf{Find a disconnected head UAV.} If a head UAV has not participated in IDBC operations for a long time and also has no response to the active request, the leader submits a UAV disconnection transaction. A new block is generated on IDBC based on the RAFT consensus protocol, and IDBC is updated, chaining the new block. The leader then looks for the information of the candidate head UAV. The head UAV disconnection transaction format is as follows.
		\begin{flalign}
			\label{equ_leaddc1}
			\scriptsize
			header = \left\{\begin{array}{lll}
				Input address = H(block)_{x0} || MerklePath_{x0}, \\ 
				Cluster index = C_x, \\ 
				Generator = UAV_{x0}.
			\end{array}\right.&&
		\end{flalign}		
		\begin{flalign}
			\label{equ_leaddc2}
			\scriptsize
			body = \left\{\begin{array}{ll}
				Type = Head\ UAV\ disconnected,& \\ 
				Payload = PK_{x0},fingerprint_{x0},T,\\
				\ \ \ \ \ Sign_{SK_{Leader}}\{H (PK_{x0},fingerprint_{x0}, T) \}.
			\end{array}\right.&&
		\end{flalign}
		
		\item \textbf{Establish connection with candidate head UAV.} According to the new block, the leader looks for the candidate head UAV and establish connection with it.
		
		\item \textbf{Generate head UAV request transaction.} After connection, the candidate head UAV generates a head UAV request transaction, whose format is as follows.
		\vspace{-0.3em}
		\begin{flalign}
			\label{equ_leaddc1}
			\scriptsize
			header = \left\{\begin{array}{lll}
				Input address = H(block)_{xk} || MerklePath_{xk}, \\ 
				Cluster index = C_x, \\ 
				Generator = UAV_{xk}.
			\end{array}\right.&&
		\end{flalign}		
		\vspace{-0.2em}
		\begin{flalign}
			\label{equ_leaddc2}
			\scriptsize
			body = \left\{\begin{array}{ll}
				Type = Head\ UAV\ update,& \\ 
				Payload = PK_{xk},fingerprint_{xk},T,\\
				\ \ \ \ \ Sign_{SK_{xk}}\{H (PK_{xk},fingerprint_{xk}, T) \}.
			\end{array}\right.&&
			\vspace{-0.3em}
		\end{flalign}
		
		\item \textls[-15]{\textbf{Consensus these transactions and generate a new block.} A new block is generated on IDBC based on the head UAV request transaction based on the RAFT consensus protocol, and IDBC is updated chaining the new block.}
		
		\item \textbf{Send and copy IDBC.} The updated IDBC is sent and copied to the new head UAV.
		
		\item \textls[+15]{\textbf{Execute key update procedure and update the white-gray list.} The new head UAV and IDBC execute the key update procedure to ensure the cluster security. In addition, the new head UAV updates the white-gray list to add the disconnected UAV.}
		
		\item \textbf{Copy task results. }\textls[-25]{After updating and broadcasting LML, First, the head UAV will collect the task results of all member UAVs in cluster $C_x$. After that, the head UAV calculates the formula \( H_2 (PK_{x0} \oplus PK_{x1}\oplus \ldots \oplus PK{xn}) \) to obtain the serial number of a member UAV, and also sends the task results initially stored in $UAV_{xk}$ to this member UAV for copying.}
	\end{enumerate}
	\begin{figure}[t]
		\centering
		\includegraphics[width=1\linewidth]{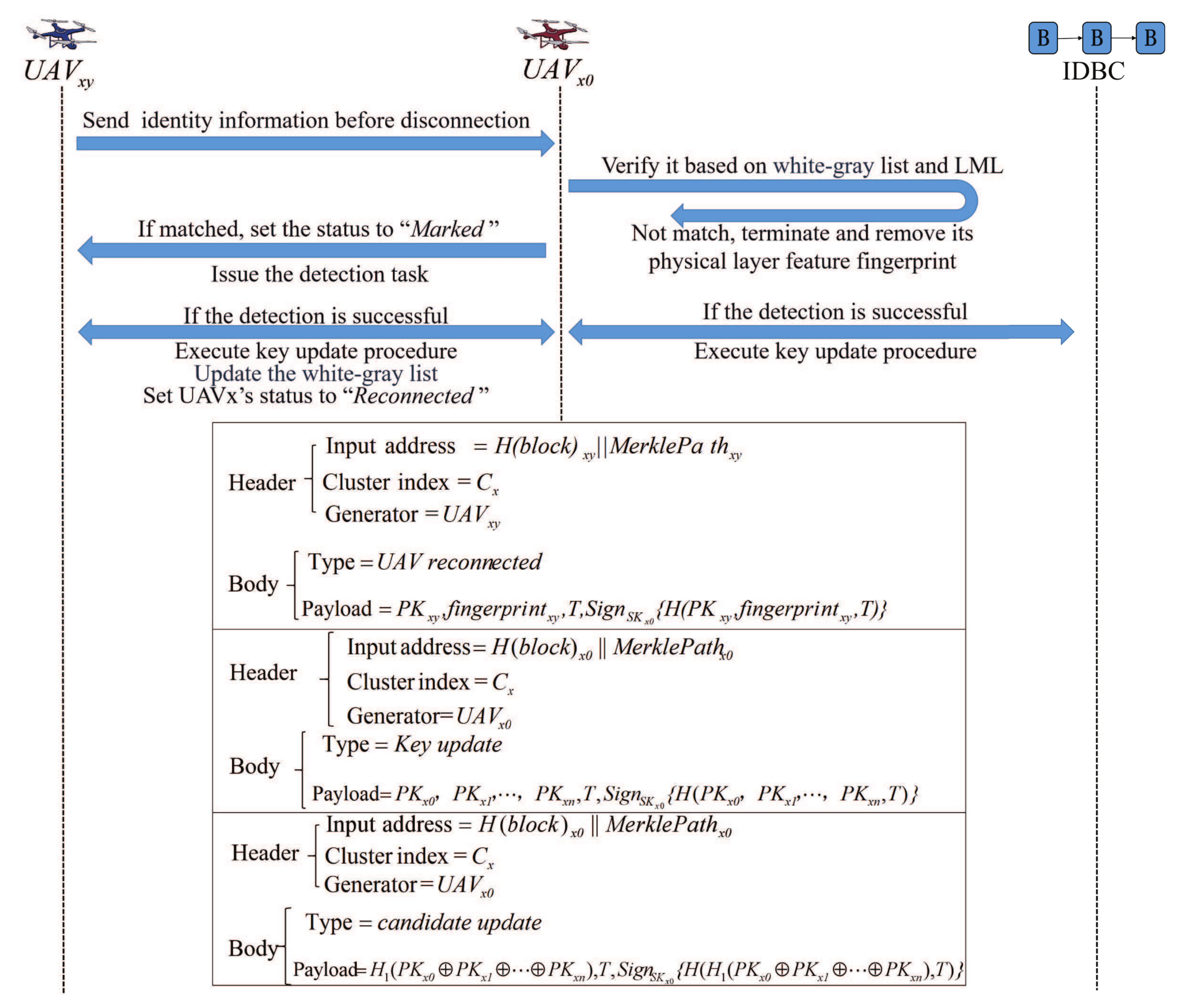}
		\caption{UAV reconnection.}
		\label{fig:UAV reconnection}
		\vspace{-0.5em}
	\end{figure}
	\vspace{-0.3em}
	\subsection {UAV Reconnection}
	\label{ssec_uavre}
	\textls[-30]{The following steps are only applicable when there is an active head UAV in the current cluster, and we assume that UAVs that disconnect will only rejoin their original clusters. Moreover, since any reconnecting UAV will ultimately be initialized as a reconnecting member UAV \cite{10272518}, the reconnection procedures for both head UAVs and member UAVs are the same. Assuming the head UAV of cluster $ C_x $ is $UAV_{x0}$, if a disconnected UAV $UAV_{xy}$ wishes to reconnect to cluster $C_x$, it should follow these steps, as shown in Fig. \ref{fig:UAV reconnection}.}
	
	\begin{enumerate}
		\item \textls[+35]{\textbf{Send identity information for reconnection.} The disconnected UAV $UAV_{xy}$ sends identity information to the head UAV for reconnection.}
		
		\item \textls[-25]{\textbf{Verify the reconnected UAV and check it.} The head UAV verifies the reconnected UAV based on the white-gray list and LML. If $UAV_{xy}$'s physical fingerprint is on the gray list and its status in LML is ``$Disconnected$", the head UAV updates the status of the reconnected UAV to ``$Marked$" in LML and assigns it a test task and executes the task together. If the results are correct, update its status to ``$Reconnected$".}
		
		\item \textls[-45]{\textbf{Execute key update procedure and update the white-gray list.} The new head UAV and IDBC execute the key update procedure to ensure the cluster security. In addition, the new head UAV updates the white-gray list to add the disconnected UAV.}
	\end{enumerate}

	\section{Experiment Evaluation}
	\label{sec_sa&ee}
	\subsection{Experiment Setup}
	\label{ssec_es}
	\textls[-15]{We deploy the experimental environment on a desktop (Intel i7-10700, CPU 2.90GHz, and 32GB RAM). The multi-cluster UAV networking is simulated using Python, and each cluster's operation area is set as a circle with a radius of $1km$. UAVs operate between altitudes of $200m$ to $1,000m$ at speeds ranging from $0-30m/s$. The head UAVs are located in the center of their cluster work area. For communication security, ECC-secp256k1 and AES128 are chosen for asymmetric and symmetric encryption with SHA256 as the hash function. We adopt the 3-D Gauss-Markov model \cite{broyles2010design} to describe the UAV's movement and an energy model \cite{tian2020blockchain} tailored for dynamic wireless sensor networks. We further set up a blockchain prototype system on Hyperledger Fabric v2.2\footnote{https://hyperledger-fabric.readthedocs.io/zh\_CN/latest/write\_first\_app.html.}, and use RAFT as the consensus protocol. The blockchain leader election period is set to every ten consensus periods.}
	
	\subsection{Robustness \& Integrity}
	
	\begin{figure}[t]
		\centering
		\includegraphics[width=1\linewidth]{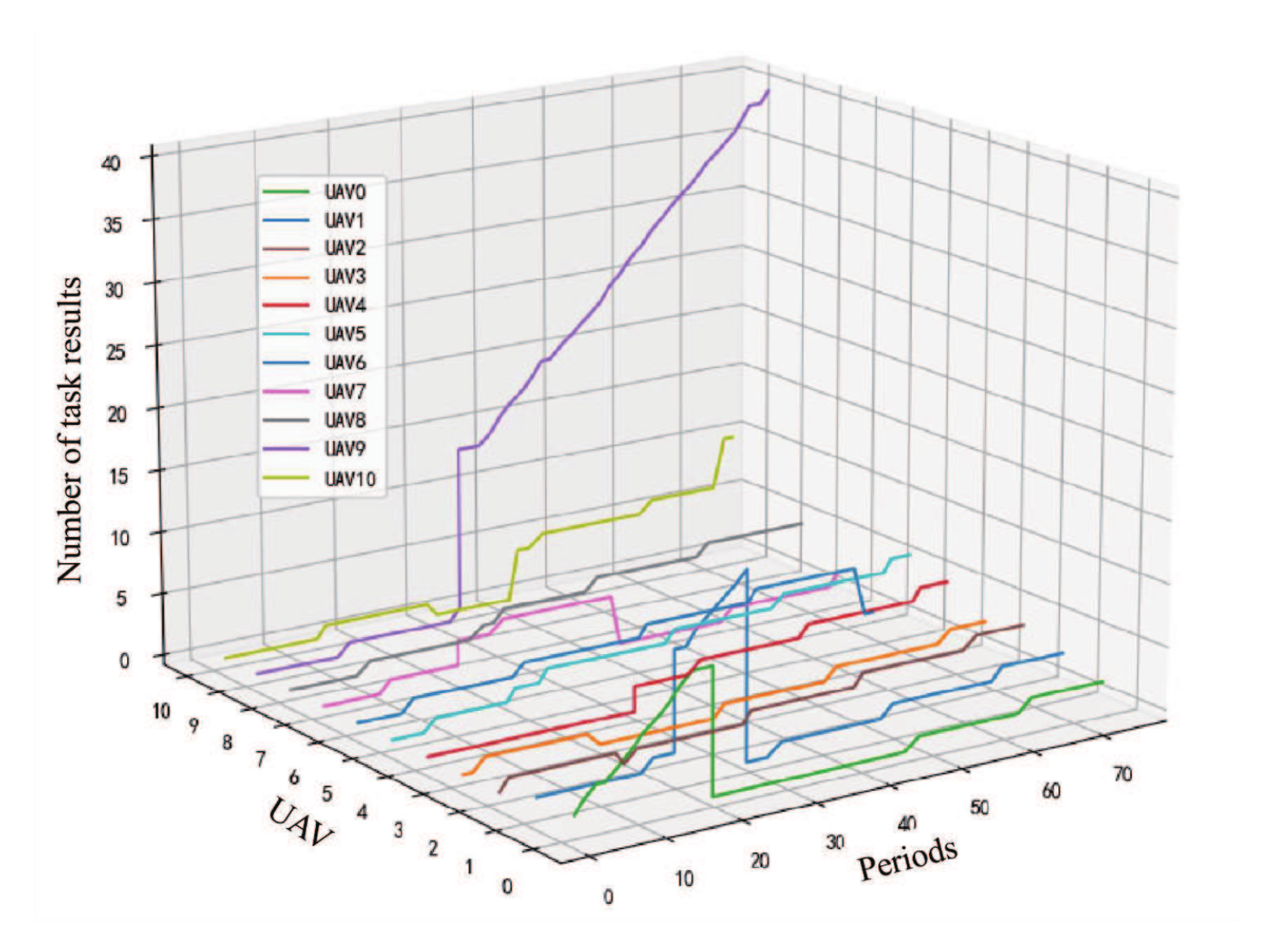}
		\caption{Robustness \& integrity.}
		\label{fig_ri}
		\vspace{-0.5em}
	\end{figure}	
	\begin{table}[t]
		\centering
		\caption{Task results stored in each UAV.}
		\small
		\setlength{\tabcolsep}{1.5mm}{
			\label{tab_tr}
			\begin{tabular}{c|c|c|c|c|c|c|c|c}
				\hline
				\diagbox{\textbf{UAV}}{\textbf{period}} &{\textbf{$1$}}&{\textbf{$2$ ... $12$}}&{\textbf{$13$}}&{\textbf{$14$}}&{\textbf{$15$ ... $47$}}&{\textbf{$48$}}&{\textbf{$49$}}&{\textbf{$50$}} \\
				\hline
				UAV$0$& $1$ &...& $10$ & $0$ &...& $2$ & $2$ & $2$ \\
				\hline
				UAV$1$& $1$ &...& $2$ & $10$ &...& $3$ & $3$ & $3$ \\
				\hline
				UAV$2$& $0$ &...& $1$ & $1$ &...& $4$ & $4$ & $4$ \\
				\hline
				UAV$3$& $0$ &...& $0$ & $0$ &...& $3$ & $3$ & $3$ \\
				\hline
				UAV$4$& $0$ &...& $0$ & $0$ &...& $5$ & $5$ & $5$ \\
				\hline
				UAV$5$& $0$ &...& $2$ & $2$ &...& $6$ & $6$ & $6$ \\
				\hline
				UAV$6$& $0$ &...& $1$ & $1$ &...& $4$ & $0$ & $0$ \\
				\hline
				UAV$7$& $0$ &...& $1$ & $3$ &...& $1$ & $1$ & $2$ \\
				\hline
				UAV$8$& $0$ &...& $1$ & $1$ &...& $5$ & $5$ & $5$ \\
				\hline
				UAV$9$& $0$ &...& $1$ & $1$ &...& $39$ & $39$ & $40$ \\
				\hline
				UAV$10$& $0$ &...& $1$ & $1$ &...& $6$ & $10$ & $10$ \\
				\hline
			\end{tabular}}
		\end{table}
		
		\textls[-15]{Fig. \ref{fig_ri} illustrates the task result storage situation of each member of a UAV cluster in a full task cycle with various events. As the task cycle goes, the sum of the number of tasks stored by UAVs in the cluster increases, and the sum of the number of tasks stored by the member UAVs is equal to that of the head UAV. Table \ref{tab_tr} lists the number of task results stored by each UAV in different periods corresponding to Fig. \ref{fig_ri} for detailed description. We initialize UAV$0$ as the head UAV, and it acts from period $1$-$13$. At the $14$-th period, the head UAV disconnects, and then the head UAV disconnection procedure is performed. Meanwhile, UAV$2$ is selected as the new head UAV, and then the task results of the cluster are copied to UAV$2$ and the original task results of UAV$2$ are copied to UAV$7$. Similarly, at the $49$-th, UAV$6$ disconnects, and its task results are copied to UAV$10$ from the head UAV. In addition, Table \ref{tab_tr} shows that the disconnected UAV can reconnect and execute tasks, like UAV$0$. It is obvious that a cluster can select a new head UAV and continue executing the rest tasks even if the head UAV disconnects, and the entirely task results are stored in the cluster through the task result backup mechanism no matter which UAV disconnects. In conclusion, the cluster robustness of our solution is apparently higher than that of the existing solution \cite{tan2020blockchain}, and our solution ensure the task result integrity.}
		
		\subsection{Energy Consumption \& Time Overhead}
		\label{ssec_ec&to}
		We assess time and energy overheads across different scenarios and analyze variances due to diverse factors, adjusting only one variable at a time in each experiment.
		
\begin{figure}[t]
			\centering
			\hspace*{-0.5cm}
			\subfloat[Comparison of miner election delay]{\includegraphics[height=1.7in]{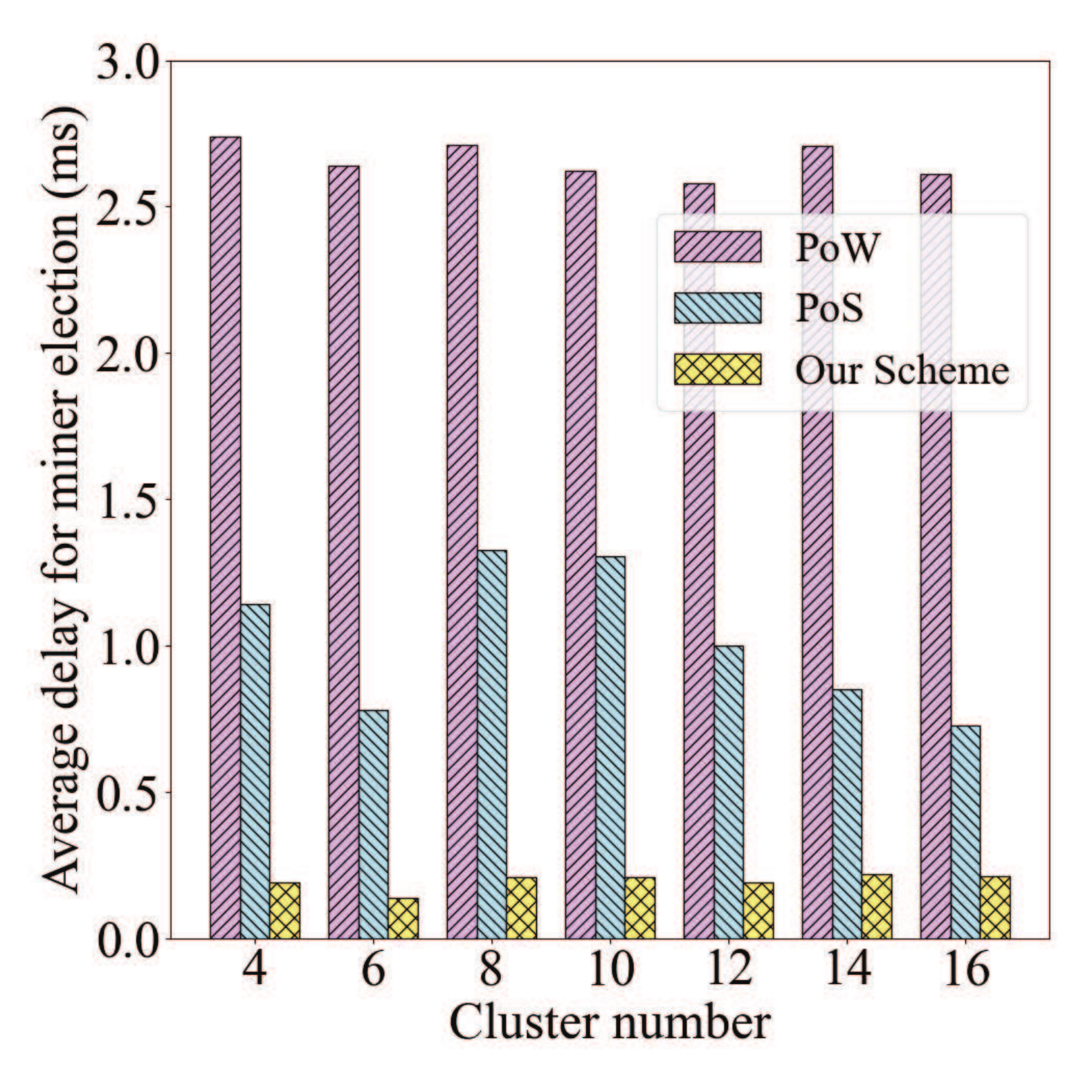}
				\label{fig_msd}}
			\subfloat[Comparison of identity authentication
			 delay]{\includegraphics[height=1.7in]{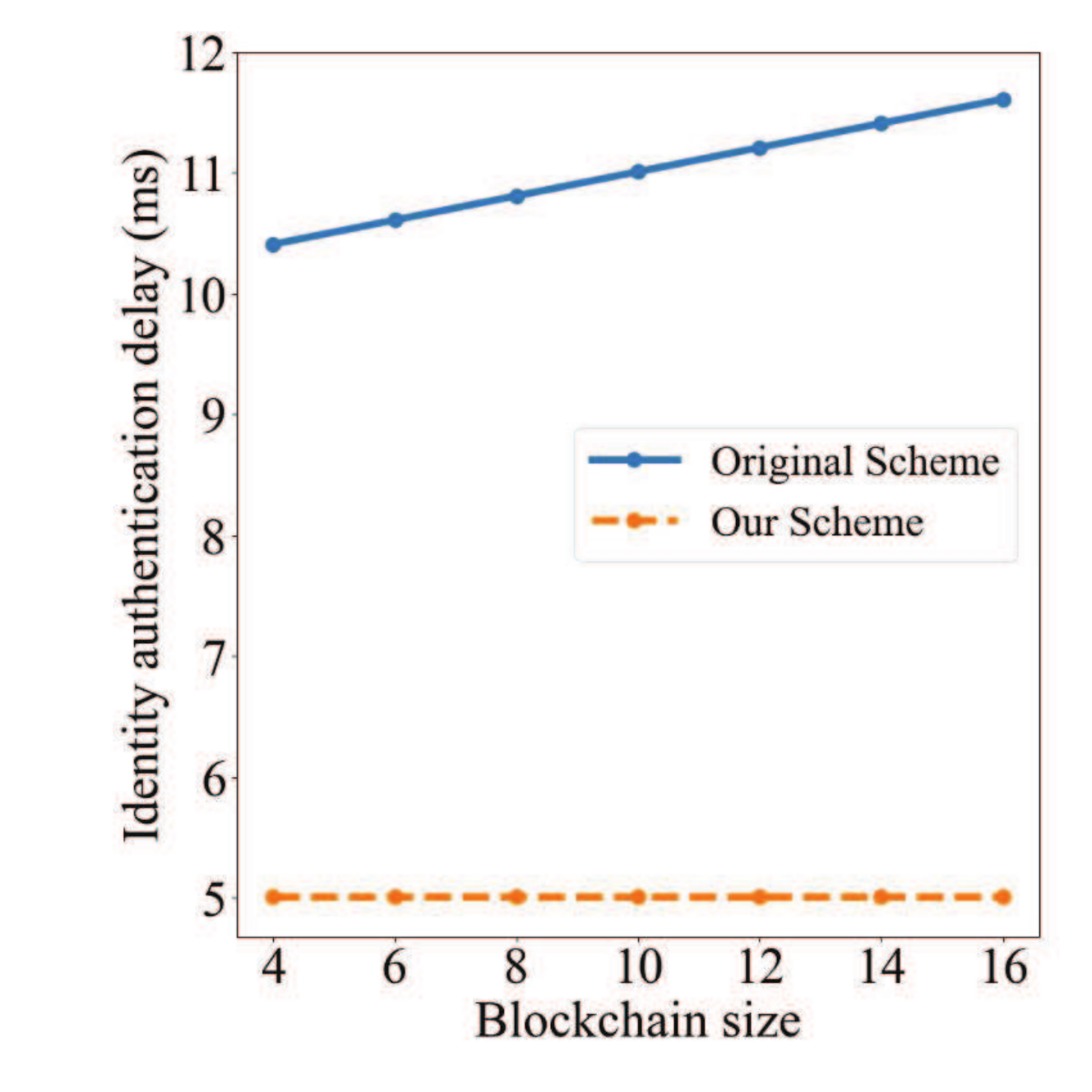}
				\label{fig_iad}}
			\caption{Delay performance.}
			\label{fig_dp}
			\vspace{-0.8em}
		\end{figure}

		\textls[-25]{Delay is a crucial indicator to evaluate the performance of blockchain prototype systems. We first discuss the effect of cluster number on the average delay of blockchain leader election with different consensus protocols. As shown in Fig. \ref{fig_msd}, the average delay of our solution is significantly lower than that of PoW and PoS and is nearly stable with the increase of cluster number. It is due to the RAFT consensus protocol we used that does not require a lot of time and power to compete with miners solving mathematical puzzles compared to the two consensus protocols. Moreover, Fig. \ref{fig_iad} illustrates that the delay of UAV identity authentication of our solution in UAV reconnection remains unchanged with the blockchain size (number of blocks), and it is obviously lower than the existing scheme \cite{tan2020blockchain}. which relies on checking a list and tracing back the blockchain to authenticate a UAV joining in another cluster after its head UAV disconnects so that the delay grows with the blockchain size. However, the identity authentication of our solution is completed by querying LML recorded on a completely trusted head UAV without querying blockchain every time. The enhanced speed is attributed to our solution's use of table lookups instead of the extensive blockchain traversal required in Ref.\cite{tan2020blockchain}.}
		
\begin{figure}[t]
			
			\centering
			\subfloat[Influence of cluster size]{\includegraphics[height=1.7in]{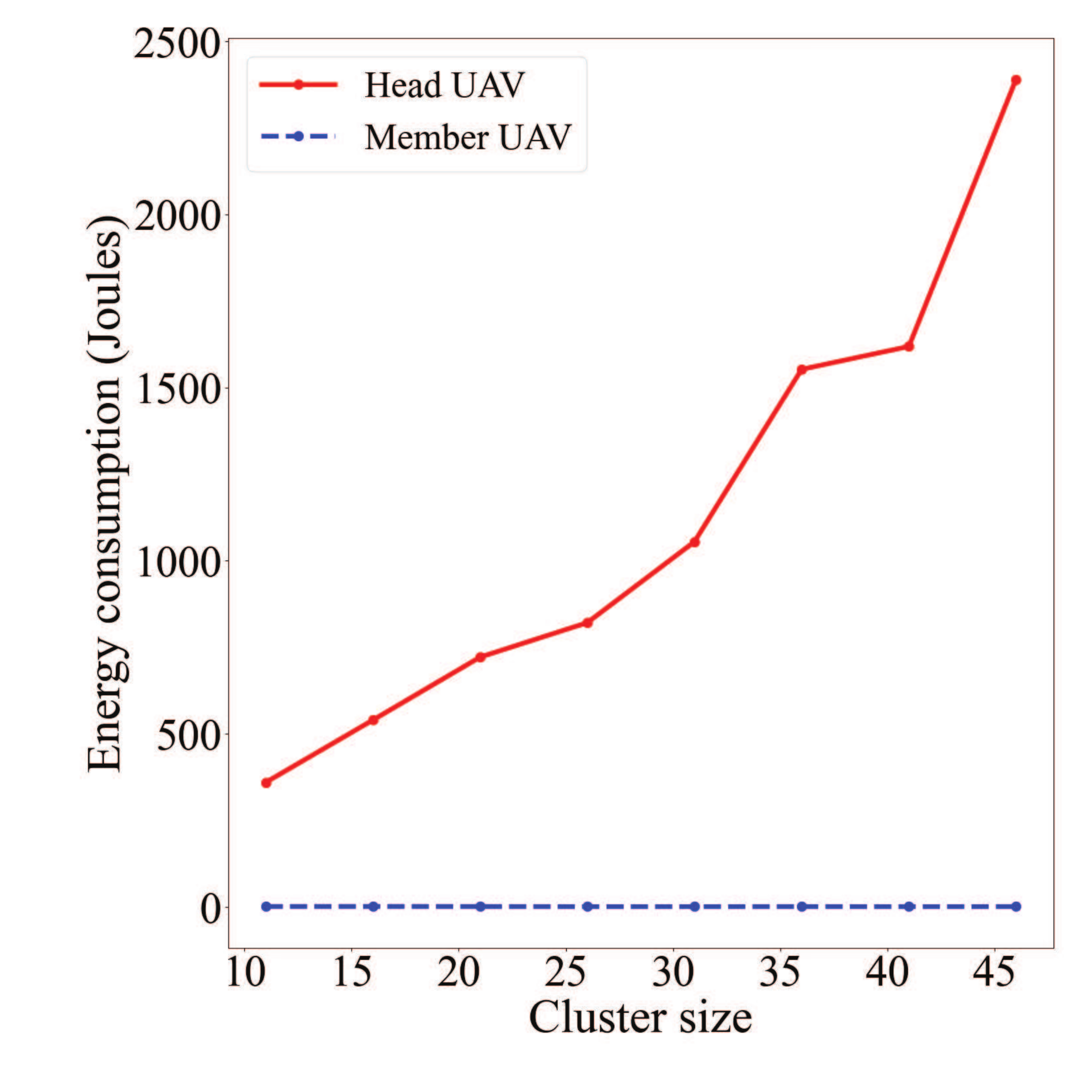}
				\label{fig_eccs2}}
			\subfloat[Influence of length of public key]{\includegraphics[height=1.7in]{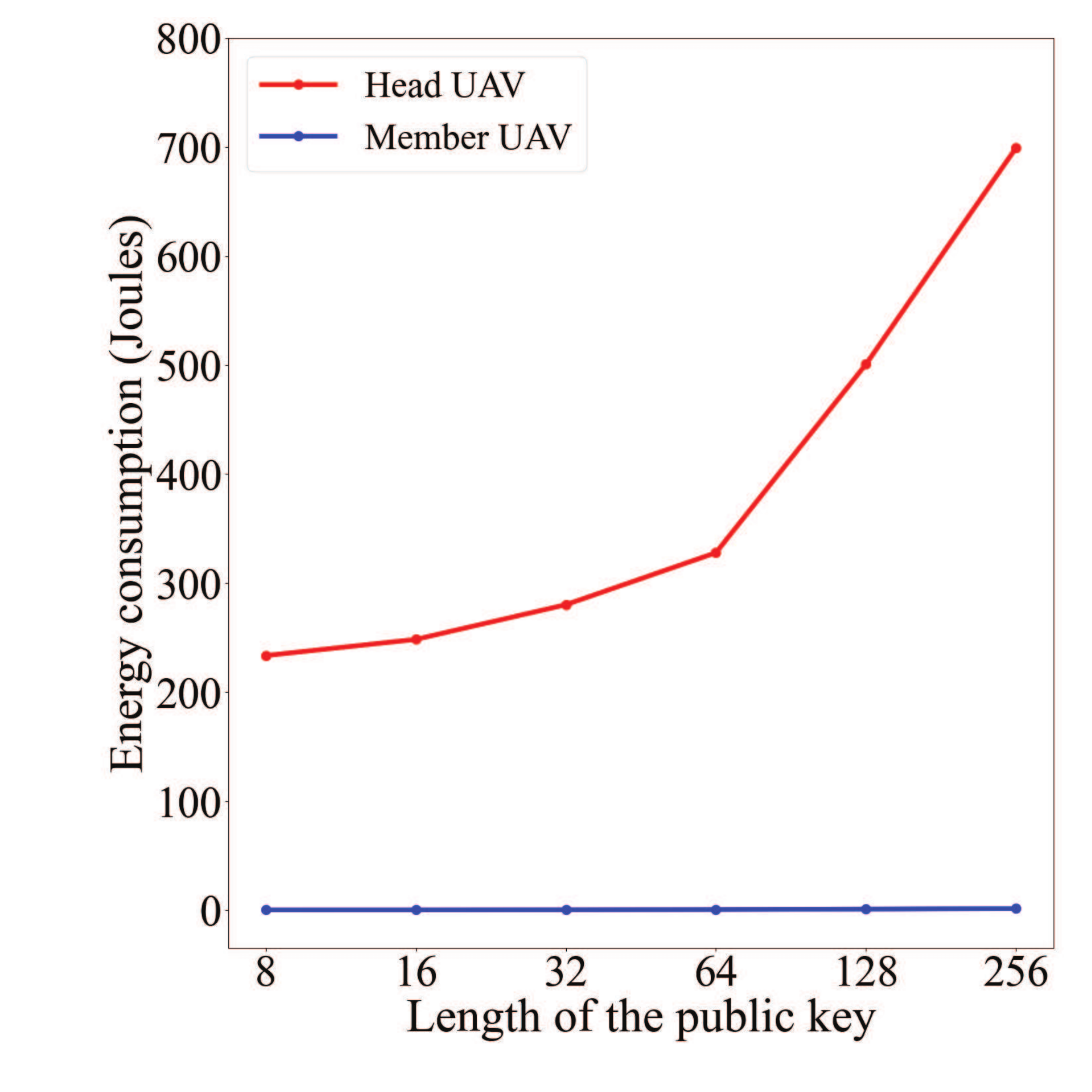}
				\label{fig_ecpkl}}
			\caption{Energy consumption of member UAV disconnection.}
			\label{fig_ecmuavdis}
			\vspace{-0.8em}
		\end{figure}

		Besides, we also analyze the energy consumption of a series of operations after the member UAV disconnection. As shown in Fig. \ref{fig_eccs2}, the energy consumption of the head UAV increases as the cluster size expands, which results in the head UAV having to process more key and member information. Meanwhile, the energy consumption of the member UAV remains unchanged because it just issues its key information and receives the updated LML. Moreover, Fig. \ref{fig_ecpkl} illustrates the energy consumption of the head UAV increases with the increase of public key length, while that of the member UAV is stable at a low value. When the public key length increases, the key data enlarges, leading to higher energy consumption.
		
		\section{Conclusion}
		\label{sec_con}
		\textls[-25]{This paper proposes a blockchain-based UAV identity authentication solution to ensure the performances of multi-cluster UAV networking operating in uninhabited areas. This solution contains a UAV reconnection mechanism and a task result backup mechanism to guarantee the cluster robustness and task result integrity. The UAV reconnection mechanism includes UAV disconnection and UAV reconnection. If UAVs or even if the head UAV disconnects, the cluster can still operate stably, and the disconnected UAVs can safely reconnect. In addition, the task result backup mechanism makes that task results are redundantly stored in both the member UAV and the head UAV, guaranteeing that the disconnection of any UAV will not result in the loss of task results. We conduct extensive experiments to demonstrate the cluster robustness and task result integrity. Experimental results also prove that our scheme's identity authentication delay and UAV energy consumption are superior compared to existing solutions, where the delay of UAV reconnection identity authentication of our solution is less than $50\%$ of that of an existing solution. In the future, we will concentrate on UAV identity authentication technology based on physical layer feature fingerprints and blockchain, without relying on the traditional PKI infrastructure, to enhance authentication efficiency.}

	\end{document}